# Mechanically Reconfigurable Terahertz Bandpass Filter Based on Double-Layered Subwavelength Metallic Rods

**Sanaz Zarei**

Sharif University of Technology, Tehran, Iran

szarei@sharif.edu

*Abstract*—Tunable bandpass terahertz filters are demanded in various key applications such as hyperspectral imagers, miniaturized spectrometers, and high-speed wireless communication systems. Here, a mechanically reconfigurable double-layered subwavelength metallic structure is presented for frequency-agile terahertz transmission bandpass filtering. The theoretically demonstrated polarization-insensitive filter shows remarkable performance metrics. By varying the vertical interlayer spacing of the metallic layers from 20μm to 4μm, the operation frequency tunes from 0.81THz to 1.32THz, and the full width at half maximum bandwidth changes from 209 GHz to 135GHz, with maximum transmission efficiency greater than 98% and quality factor ranging between 3.88 and 9.77. A larger variation range of the vertical interlayer spacing leads to an enhanced frequency tuning range of the filter. Furthermore, simultaneous vertical and lateral interlayer displacements can provide polarization-dependent behavior for the filter. The underlying physical mechanism governing the filter's frequency response and tuning capability is analyzed by examining the electromagnetic field distributions within the double-layered subwavelength metallic structure and its Fabry–Pérot-like behavior. The presented scheme holds significant promise for many terahertz applications due to its large tuning range, easy tuning mechanism, simple structure, and compatibility with fabrication materials and processes.

*Keywords*—terahertz filter, mechanical tuning, subwavelength metallic structures, metamaterials

## 1. Introduction

The terahertz frequency spectrum, spanning from 0.1–10THz, holds significant promise for a wide range of disruptive applications such as the next-generation high-speed 6G wireless communication, medical imaging, spectroscopy, and sensing, enabled by its broad available bandwidth, ability to penetrate through many optically opaque materials, (sub-)millimeter spatial resolution, and the occurrence of unique rotations, vibrations or liberations of molecules and molecular aggregates within this frequency range [1]. Nevertheless, because of the difficulties in scaling conventional microwave electronics and infrared photonic technologies to terahertz frequencies, and the weak interaction of terahertz waves with natural materials, the realization of efficient and compact terahertz devices becomes complicated, which has hindered the rapid progress of the terahertz technology [2]. To address these challenges, recent focus has turned to engineered artificial materials called metamaterials [3]. Metamaterials consist of arrays of subwavelength unit cells, whose electromagnetic responses can be engineered by adjusting the unit cells' geometry and dimensions [2]. Made from subwavelength metal and dielectric constituents, they provide an effective way to manipulate terahertz waves [4]. Numerous terahertz functional devices for manipulating the amplitude, phase, frequency, and polarization of terahertz waves have been demonstrated, exploiting the outstanding electromagnetic properties of metamaterials. Dynamic tunability of metamaterials' properties can be achieved, either by integrating active materials like liquid crystals, ferroelectric or ferromagnetic materials, graphene, and phase change materials into their unit cells or by mechanically reconfiguring their unit cells employing microelectromechanical systems (MEMS)-based actuation mechanisms [2, 4-5]. The former suffers from a limited modulation range, whereas the latter relies heavily on a complex structure that presents significant fabrication challenges [4].

As one of the most important terahertz functional devices, terahertz bandpass filters play a vital role in terahertz imaging and high-speed wireless communication systems [6]. They selectively transmit the terahertz light at specific frequencies and effectively eliminate the unwanted components [7]. Tunable transmission frequency is one of the most prominent features of a terahertz filter. However, the realization of spectrally tunable terahertz filters is still challenging, primarily because of the lack of natural materials exhibiting substantial tunable refractive indices at terahertz frequencies, as well as large device footprint and the fabrication complexities associated with metallic grid lines and resonators [2]. Therefore, the design of a continuously tunable terahertz bandpass filter that offers broad frequency tuning range and high spectral selectivity is immensely demanded [8].

The paramount effect of surface plasmons in tunable filters has already been demonstrated [9]. The extraordinary optical transmission (EOT) through subwavelength metallic structures, driven by the resonance excitation of surface plasmons, enables effective filtering and frequency tuning in these structures. By taking advantage of the frequency-selective resonance properties of subwavelength metallic structures, various terahertz functional devices such as filters, modulators, and switches can be realized [9]. The geometric configuration of the subwavelength metallic structures exerts a substantial influence on their terahertz transmission behavior [10], underscoring the importance of a reasonable structural design. Many previous works studied the effect of metal film thickness, periodicity, aperture geometry, etc., on the properties of extraordinary terahertz transmission [11]. Multi-layer stacking of subwavelength metallic structures can also provide desirable transmission characteristics [8, 10, 12-18] and enable tailoring their resonance properties without modifying individual unit cells, achieved instead by varying the arrangements of layers against each other [8, 13-18]. This article proposes a tunable bandpass terahertz filter with the working frequency spanning from 0.81THz to 1.32THz. The spectral tunability of the filter is achieved by mechanically varying the vertical spacing between a pair of identical subwavelength metallic rod arrays, each of which is attached to a supporting substrate. Due to the free-space nature of the filter, it is designed to be insensitive to the polarizations of impinging waves [8]. The impact of lateral misalignment between the layers is also considered.

## 2. PROPOSED STRUCTURE AND ITS PERFORMANCE

A unit cell of the reconfigurable terahertz bandpass filter based on the double-layered subwavelength metallic rod array is illustrated in Fig. 1(a). The metallic rods in each array are fixed to a polystyrene foam substrate. The two metallic layers are vertically movable relative to one another, allowing their vertical interlayer spacing to be variable. The upper substrate holding the metallic rods is mechanically coupled to a MEMS actuator that facilitates precise vertical displacement. The metallic rods that are fixed to the lower substrate are aligned center-to-center with the upper metallic rods. Upon application of a voltage, the MEMS actuator induces a vertical motion of the suspended substrate, changing the vertical interlayer spacing of the two metallic layers. Key geometric parameters of the presented filter include the lattice constant ($\Lambda$ = 100µm), the radius of the metallic rods ($R_r$ = 47.5µm), the thickness of the metallic rods ($H_1$ = 20µm), the thickness of the supporting substrates ($H_2$ = 100µm), and the interlayer vertical spacing between the structured metallic layers (4µm < Spacing < 20µm). These values were selected to ensure ease of fabrication and measurement while maintaining the desired electromagnetic performance.

Electromagnetic simulations were conducted using a commercial finite element method (FEM) solver. To minimize the computation time, the model was simplified to a single unit cell, incorporating periodic boundary conditions to emulate an infinitely periodic array. The metallic rods were modeled as perfect electric conductors (PECs), an acceptable approximation at terahertz frequencies due to the negligible metallic losses. Due to the fourfold rotational symmetry of the design, the transmission characteristics are invariant under orthogonal polarizations. Figure 1(b) presents the transmission spectrum of terahertz waves through the filter. When the interlayer vertical spacing is 20µm, the transmission peak is located at 812GHz with a bandwidth of approximately 209GHz. This results in a quality factor of 3.88. Actuation of the MEMS device causes the vertical displacement of the upper layer, reducing the interlayer vertical spacing. Thereby, the transmission peak blueshifts to higher frequencies, and at the vertical spacing of 4µm, the transmission peak gets to 1.32THz with a full width at half maximum bandwidth of 135GHz, bringing about a quality factor of 9.77. The maximum transmission efficiency is above 98% through the whole tuning range. For the presented results in Fig. 1(b), the supporting substrates were excluded in the simulations.

Figure 2 illustrates the electromagnetic fields distributions through the filter for two representative interlayer vertical spacings of 4µm and 20µm, at two frequencies of 812GHz (corresponding to the resonance frequency at 20µm interlayer spacing) and 1.32THz (corresponding to the resonance frequency at 4µm interlayer spacing). It can be deduced from these sketches that the coupling between the evanescent fields of the two layers is enhanced by decreasing the interlayer distance of the layers. This can be implied by comparing the magnitude of the magnetic fields for the interlayer spacing of 4µm with those of the interlayer spacing of 20µm. Also, it can be inferred that for each interlayer vertical spacing, at the resonance frequency (Figures 2(b) and 2(c)), there is a strong coupling between the evanescent fields of the layers inside the interlayer spacing, which leads the electromagnetic fields to flow between the walls of the neighboring rods in the second metallic layer and then flow out. In contrast, at frequencies outside the resonance bandwidth (Figures 2(a) and 2(d)), the coupling between the evanescent fields of the layers through the interlayer spacing is so weak that most of the electromagnetic waves cannot be guided inside the second layer, and they are reflected by the first layer. It is noteworthy that a single layer of similar subwavelength metallic rod array exhibits relatively low terahertz transmittance across the entire range of 812GHz to 1.32THz (that is the designated tuning range of the filter). Within this spectral range, the majority of incident terahertz radiation is reflected by the single-layer structure (please refer to Fig. 4). Consequently, the transmission peaks appeared in the filter's frequency response are mainly generated by the strong evanescent field coupling between the two stacked layers of subwavelength metallic structures.

The employed substrate material for this structure is polystyrene foam that has an exceptionally low refractive index (ranging from 1.017 to 1.022) and minimal absorption losses within the terahertz frequency range of 0.2–4THz [19]. For including the supporting substrates in the simulations, two approaches can be followed:

a) The complete geometry of the filter, including the foam substrates and the extended height of the rods pinned inside the substrates, is incorporated in the simulation model. The extended height of the rods inside the substrates is 10µm. Figure 3 presents the simulation results for this configuration. By changing the interlayer vertical spacing from 20µm to 4µm, the transmission peak tunes from 703GHz to 1.16THz. The existing deviations in resonance frequencies between this configuration and the one excluding the supporting substrates (Fig. 1(b)) are due to the 10µm difference in the height of the rods (30µm for this configuration versus 20µm for the one presented in Fig. 1(b)).

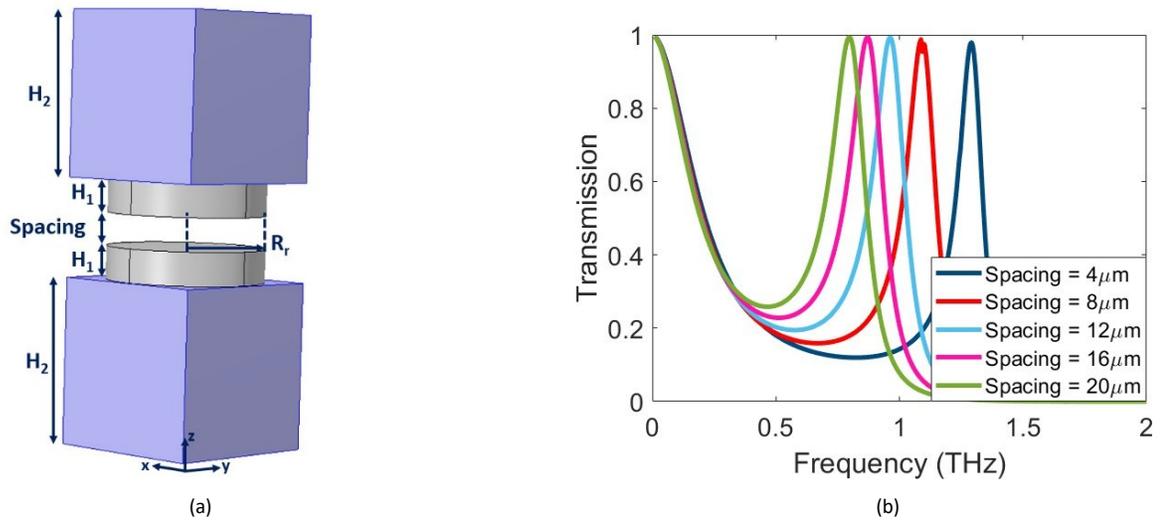

**Fig. 1** (a) Schematic of a unit cell for the tunable terahertz filter. By changing the vertical spacing between the two metallic circular rods that are fixed to the substrates, the transmission passband frequency of the filter can be adjusted. (b) The transmission characteristics of the filter for incident terahertz light of any linear polarization. The dark blue, red, light blue, pink, and green curves show the filter transmission for vertical spacings of 4μm, 8μm, 12μm, 16μm, and 20μm, respectively. For calculating the transmission diagrams in (b), the supporting substrates were excluded in the simulations.

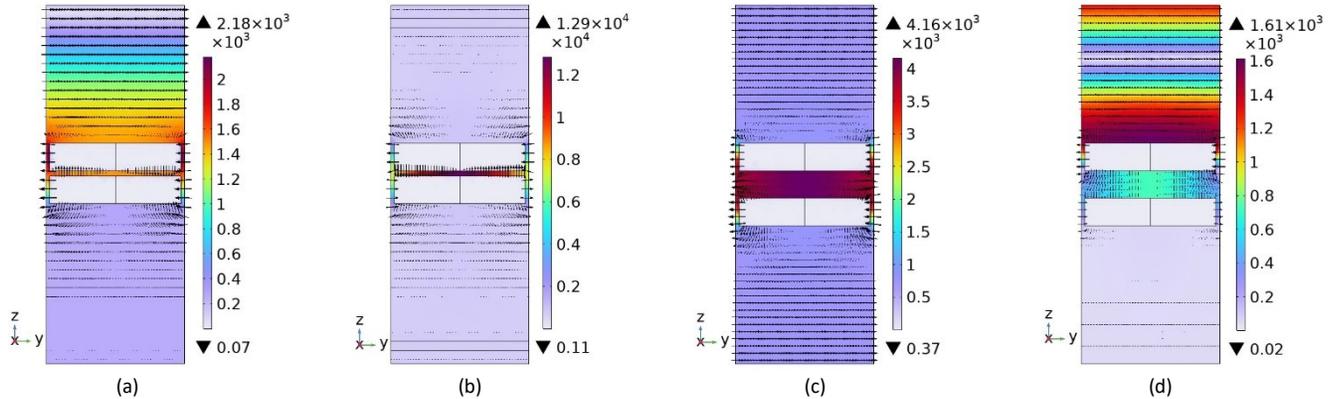

**Fig. 2** Simulated electromagnetic field distribution throughout the terahertz filter for (a) spacing of 4μm at 0.81THz, (b) spacing of 4μm at 1.32THz, (c) spacing of 20μm at 0.81THz, (d) spacing of 20μm at 1.32THz. The 0.81THz and 1.32THz correspond to the peak transmission frequencies at 20μm and 4μm, respectively. The magnetic fields are aligned along the x-axis, and their magnitudes are represented by the color bars, while the electric fields are shown by arrows (without scale).

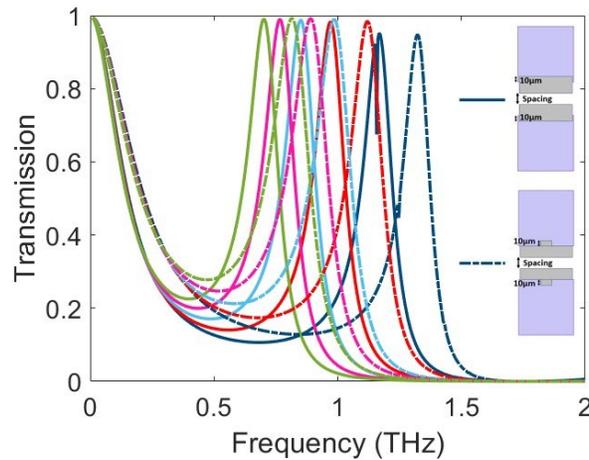

**Fig. 3** The transmission characteristics of the designed terahertz filter for incident terahertz light, considering the supporting substrates in the simulations. There are two options for fixing the metallic circular rods to the substrates, one by anchoring the rods into the substrate with the depth of 10μm within the substrate (solid lines), and the other by making mushroom-like rods and anchoring them within the substrate from the pin-like side (dotted-dash lines). The pins' height within the substrate is 10μm. The dark blue, red, light blue, pink, and green curves show the filter transmission for vertical spacings of 4μm, 8μm, 12μm, 16μm, and 20μm, respectively.

b) The complete geometry of the filter, including the foam substrates and the extended height of the rods pinned inside the substrates, is incorporated in the simulation model; however, the extended height of the rods has a reduced radius (10µm instead of 47.5µm). The extended height of the rods that is pinned inside the substrates is also measured as 10µm. As can be observed in Fig. 3, for this configuration, the tuning range spans from 812GHz for 20µm interlayer vertical spacing to 1.32THz for 4µm interlayer vertical spacing, which perfectly resembles the tuning performance of the configuration without the supporting substrates (Fig. 1(b)).

## 3. Discussion

Surface plasmon polaritons (SPPs) are electromagnetic surface waves confined to the metal–dielectric interface. At terahertz and microwave frequencies, where metals behave like perfect electric conductors, "spoof" or "designer" SPPs can be realized by introducing subwavelength surface features—such as periodic grooves or apertures—that enable enhanced field penetration and tunable plasmonic dispersion [20]. In parallel, localized surface plasmons (LSPs) arise at the edges of subwavelength apertures, supporting highly confined modes that couple with SPPs to direct their propagation and facilitate enhanced transmission. The combined interaction between SPPs and LSPs underlies the mechanism of extraordinary optical transmission (EOT), with resonance characteristics determined by geometric parameters like aperture size, spacing, periodicity, and the surrounding dielectric environment [21].

In periodically structured surfaces with metallic rods, spoof SPPs propagate as guided modes, with each adjacent rod pair functioning as an effective transmission line [20]. The interaction between the dominant fields within the rod gaps and the surface plasmons of the rods is analogous to that in periodically corrugated surfaces featuring a one-dimensional (1D) array of slits. Such corrugated surfaces support TM-polarized spoof SPPs. Similarly, spoof SPPs on textured surfaces with rods exhibit TM-like surface wave characteristics, and their dispersion curves closely match those of the 1D slit arrays [20]. This analogy is particularly evident for periodically structured surfaces composed of metallic square rods. A textured metallic surface with square rods (forming a two-dimensional grid) differs from a 1D grating (a periodically corrugated surface with a 1D array of slits) only by the presence of additional periodic gaps (slits) in the strips of the grating, which are oriented parallel to the electric field and surface currents in the strips, and therefore, their presence or absence has minimal influence on the surface currents, and this effect must completely vanish as the gap width approaches zero [22]. Also, the dependence of the characteristics of spoof SPPs on the rods' shape, in textured metal surfaces with square arrays of metallic rods, is previously investigated in [20] using a full-vector finite element method (FEM). It was shown that the dispersion of spoof SPPs in textured surfaces with rods is independent of the rods' shape. Here, we consider three kinds of textured metal surfaces: a square array of subwavelength metallic circular rods, a square array of subwavelength metallic square rods, and a 1D array of subwavelength metallic slits. To perform a good comparison, all three metallic gratings have the same thickness of $H_1$ = 20µm and the same lattice constant of $\Lambda$ = 100µm. The area cross-section of the rods in 2D gratings is similarly equal to $\pi \times R_r \times R_r = \pi \times (47.5\mu m)^2$. This leads to a rod width of 84.2µm in the square array of subwavelength metallic square rods. Analogously, in the 1D array of subwavelength metallic slits, the strip width is set to 84.2µm. Figure 4(a) shows the transmission characteristics of each metallic grating in the range of 0.01-4THz. The transmission spectrum

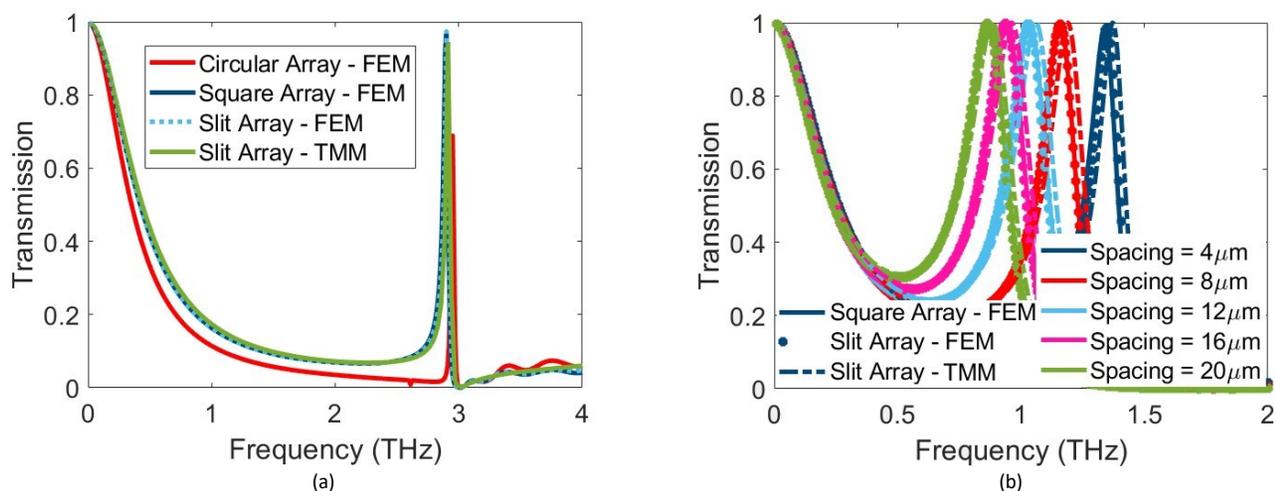

**Fig. 4** (a) The transmission characteristics of a one-layer metallic circular rod array (red curve), metallic square rod array with the same rods' cross-sectional area (dark blue curve), and metallic slit array (1D analogous of the metallic square rod array) that is calculated by both FEM (light blue dotted curve) and TMM (green curve). (b) The transmission characteristics of a double-layered metallic square rod array with the same rods' cross-sectional area as the double-layered metallic circular rod array of Fig. 1(b) (solid line), and a double-layered metallic slit array (1D analogous of the double-layered metallic square rod array) that is calculated by both FEM (star marks) and TMM (dashed line). The dark blue, red, light blue, pink, and green curves show the filter transmission for vertical spacings of 4µm, 8µm, 12µm, 16µm, and 20µm, respectively.

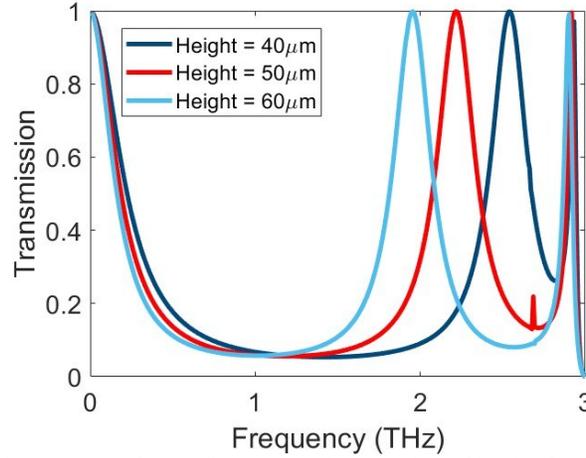

**Fig. 5** The transmission characteristics of a metallic circular rod array with rods' height of 40µm (dark blue curve), 50µm (red curve), and 60µm (light blue curve), showing the dependence of transmission frequency on the rods' height.

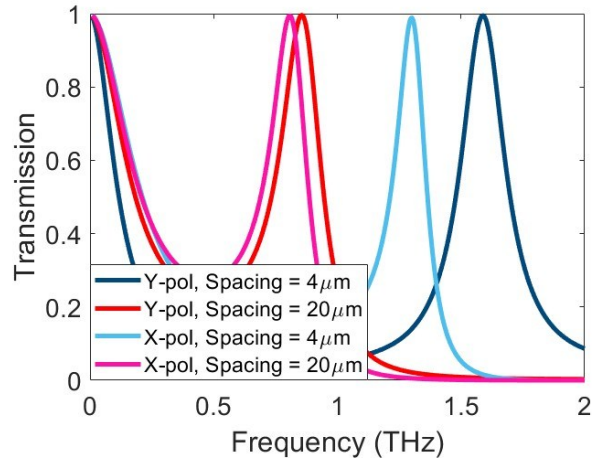

**Fig. 6** The transmission characteristics of the terahertz bandpass filter, when there is 50µm lateral misalignment between the metallic layers along the y-axis. The dark blue and red curves indicate the transmission for y-polarized incident terahertz light, when the interlayer spacing is 4µm and 20µm, respectively. The light blue and pink curves indicate the transmission for x-polarized incident terahertz light, when the interlayer spacing is 4µm and 20µm, respectively.

of the 2D array of square rods is closely matched to that of the 1D array of slits, as is indicated in [22]. The transmission spectrum of the 2D array of circular rods slightly deviates from those of the other two gratings. The transmission spectrum of the 1D array of slits that is calculated by FEM is further recalculated by the transfer matrix method (TMM) [17, 23] for verification. The simulation results presented in Fig. 4(a) indicate that the transmission spectrum of textured metal surfaces with square arrays of rods is almost independent of the rods' shape. The same conclusion holds for the double-layered gratings of the same corrugations. Therefore, the tuning performance of the presented terahertz filter should be maintained if circular rods are replaced with square rods of the same cross-sectional area (Fig. 4(b)). Also, for the purpose of validation, the tuning performance of a filter based on 1D arrays of slits, calculated by FEM and TMM, is also presented in Fig. 4(b).

In addition, the transmission spectrum of a single-layered metallic grating composed of subwavelength metallic circular rods is investigated by changing the grating thickness from 40µm to 60µm. As can be observed in Fig. 5, by increasing the grating thickness, the resonance frequency of the grating reduces from 2.54THz to 1.95THz. This trend in resonance frequency change resembles the one for the presented tunable filter by increasing the interlayer spacing between the two layers. Therefore, increasing the interlayer spacing between the two aligned identical 2D metallic gratings produces the same effect as increasing the thickness of an individual 2D metallic grating. Furthermore, a 2D metallic grating with 40µm thickness (Fig. 5) can, in fact, be equivalent to a double-layered 2D metallic grating without interlayer vertical spacing (spacing = 0), that is, two 20µm-thick 2D metallic gratings which are completely tangent together [24]. This interpretation provides insight into the situation where the interlayer spacing between the two metallic layers becomes even less than 4µm (Fig. 1(b)), until the two layers closely stick together. It can be deduced from Fig. 5 that by decreasing the interlayer spacing to zero, the resonance frequency tunes to 2.54THz that is much larger than the resonance frequency reported for 4µm interlayer spacing. Therefore, the frequency tuning range of the terahertz bandpass filter can be further enlarged by increasing the variation range of the interlayer spacing.

The physical mechanism behind the tuning behavior of the presented filter can also be explained by considering the double-layered subwavelength metallic rod array as a Fabry-Perot resonance cavity, in which each subwavelength metallic grating acts as a Fabry-Perot wall and by constantly changing the distance between the Fabry-Perot walls, the resonance frequency of the cavity can be continuously altered.

Although the presented filter is polarization-insensitive, it is possible to get polarization-dependent performance from it by creating a lateral interlayer displacement between the layers (in addition to the vertical interlayer displacement). Figure 6 depicts the performance of the filter for two orthogonal polarizations when there is a 50µm lateral misalignment between the layers along the y-axis. It is evident that the tuning range for y-polarization extends from 857GHz to 1.59THz, which is far beyond the tuning range of the filter without lateral misalignment, while for x-polarization, no variation happens in the tuning range of the filter. In an earlier work, the lateral displacement between two identical layers of subwavelength metallic slit arrays was exploited to realize a terahertz frequency-tunable filter [17]. The lateral displacement between identical layers of subwavelength metallic rods can also be applied for frequency tuning.

## 4. Conclusion

In summary, this article proposes a terahertz tunable passband filter based on double-layered subwavelength metallic rod arrays, offering a wide reconfigurable frequency range from 812 GHz to 1.32 THz with a peak transmission efficiency exceeding 98%. The filter's full-width at half-maximum (FWHM) bandwidth varies between 209 GHz and 135 GHz across the tuning range. Reconfigurability is achieved by adjusting the interlayer vertical spacing between the two metallic layers from 20 µm to 4 µm. The tuning range can be extended by more than three times, if the variation range of the interlayer vertical spacing can be increased. While the structure is inherently insensitive to the polarization of incident terahertz radiation, it can exhibit polarization-dependent frequency responses when lateral misalignment is introduced between the layers.

## STATEMENTS & DECLARATIONS

### FUNDING

The author declares that no funds, grants, or other support were received during the preparation of this manuscript.

### COMPETING INTERESTS

The authors have no relevant financial or non-financial interests to disclose.

### AUTHOR CONTRIBUTIONS

S. Z. performed the study conception and design, accomplished the FEM simulations, analyzed the results, and wrote the manuscript.

### DATA AVAILABILITY

No datasets were generated or analyzed during this study.

### ETHICS DECLARATIONS

This article does not contain any studies involving animals or human participants.